\documentclass{aa}  
\usepackage{graphicx}
\usepackage{txfonts}
\usepackage{aalongtable}
\usepackage{longtable}

\usepackage{natbib}
\bibpunct{(}{)}{;}{a}{}{,}

\input epsf.sty

\newcommand{\KU}{\mbox{\rm 4U 1630-472}}

\begin{document}
   \title{Disk emission and atmospheric absorption lines in 
 black hole candidate 4U 1630-472 }

   \author{A. R\'o\.za\'nska \inst{1}
	  \and
          J. Madej \inst{2}
          \and 
          P. Bagi\'nska \inst{1}
          \and 
          K. Hryniewicz \inst{1}
          \and         
          B. Handzlik \inst{2}
          }
   \offprints{A. R\'o\.za\'nska}

   \institute{  N. Copernicus Astronomical Center,
		Bartycka 18, 00-716 Warsaw, Poland \\
		\email{agata@camk.edu.pl}
                \and
                Warsaw University Observatory, 
                Al. Ujazdowskie 4,
                00-478 Warsaw, Poland \\
                \email{jm@astrouw.edu.pl}		
             }

   \date{Received ????, .....; accepted ???, .....}


 
  \abstract
   {We re-analyzed SUZAKU data of the black hole candidate
     \KU\ being in the high/soft state. The source, known for X-ray 
    outbursts and for absorption dips, has X-ray continuum spectrum  
    well interpreted as an accretion disk emission.
   Additionally, two absorption lines from He and H-like iron were clearly 
  detected  in \KU\ high resolution X-ray data.
   }
   {We show that the continuum X-ray spectrum of \KU\ 
 with iron absorption lines can be satisfactorily modeled by the 
   spectrum from an accretion disk atmosphere.
  Absorption lines of highly ionized iron originating in hot 
accretion disk atmosphere can be an alternative or complementary
 explanation to the wind model usually  favored for these type of sources.
  }
   {We perform full radiative transfer calculations to model 
the emission from an accretion disk surface which is seen at 
different viewing angles using our transfer code ATM21. Computed models 
are then fitted to  high resolution X-ray spectra of \KU\ obtained by   SUZAKU
 satellite.
}
   {We model continuum and line spectra using a  single model. 
Absorption lines of highly ionized iron can origin 
in  upper parts of the disk atmosphere which is intrinsically hot due 
to high disk temperature. Iron line profiles computed with 
natural, thermal and pressure broadenings match very well observations.}
   {According to any global disk models considered for the mass of 
central object of the  order 10 $M_{\odot}$ or less, the effective
 temperature of the inner radii reaches $10^7$K. 
 We showed that the accretion disk atmosphere can effectively produce iron
 absorption lines observed in \KU\ spectrum. 
Absorption line arising in accretion disk atmosphere is 
 the  important part of the observed line profile, 
 even if there are also other  mechanisms responsible for the absorption 
  features. Nevertheless, the wind theory can be an artifact of the fitting
 procedure, when  the continuum and lines are fitted as separate 
  model components.
}

   \keywords{X-rays: binaries, Stars: individual, Accretion, accretion disks, 
          Radiative transfer, Line: profiles, }

\maketitle
%

\section{Introduction}

Emission from the accretion disk around a compact object is the 
commonly accepted model for the soft X-ray bump observed in 
Galactic Black Hole candidates (GBHc). 
Depending on the mass of the central object and on the accretion rate, 
the inner disk temperature for the black hole of 10 $M_{\odot}$
can be of the order of $ 10^7$ K \citep{rozanska2011}. 
According to the accretion disk theories, either geometrically thin, optically
 thick standard 
disk \citep[][hereafeter SS disk]{SS73}, even with relativistic  
corrections \citep[][hereafeter NT disk]{novikov73},  or slim disk 
\citep{abramowicz88}, 
temperature always increases with 
decreasing mass of the central object and increasing accretion rate.

For such high temperatures of an accretion disk atmosphere, thermal lines 
form H-like and He-like iron ions should be formed, and  
those lines should be visible between 6.7 to 9.2 keV, 
where the latter value is energy of the last iron bound-free transition 
(from H-like ion to complete ionization).
Nevertheless, it is well known that observed X-ray spectra from 
compact binaries not always exhibit disk component. 
Due to instrument's technical limits and due to 
spectral state transition, multi-temperature disk component 
may not be detected in full X-ray energy range.
To explain data by the accretion disk emission we have to be sure 
that observations were taken when the source was in so called 
``soft state'' - dominated by disk-like component.  
 
Recently, several X-ray binaries do exhibit absorption 
lines from highly ionized iron \citep{boirin2004,kubota2007,trigo12}.
Many of those sources show dips in their light curves, which are believed
to be caused by obscuration of the central X-ray source by a dense material 
located at the outer edge of an accretion disk. 
Such obscuring material was accumulated during accretion phase from the 
companion star onto the disk \citep{white82}. The presence of dips and
lack of total X-ray eclipses by the companion star indicate that the 
system is viewed relatively close to edge-on, at an inclination angle in 
the range $\sim 60-80^{\circ} $ 
\citep{frank87}.

He-like and H-like Fe  absorption features indicate that highly ionized
plasma is present in these systems. Study of these lines is extremely
important to characterize the geometry and physical properties of plasma. 
Recently, it was shown that the presence of absorption lines is not 
necessarily related to the viewing 
angle since non-dipping sources also show those features \citep{trigo12}. 
Moreover, the Fe{\sc xxv} absorption line was also observed during  
non-dipping intervals in XB1916-053 \citep{boirin2004}. 

Usual way of modeling highly ionized Fe absorption lines from those sources
at first is to fit continuum to the spectrum, given either by 
multi-blackbody disk emission or by powerlaw shape. 
After fitting of the continuum  
each line is fitted separately using Gaussian shape. 
In such a way, equivalent widths (EWs) and velocity shifts of the 
line centroid can be derived. In most cases absorbing material is outflowing 
with velocities reaching few hundred km/s. The extreme example is the  
recent observation of black hole candidate IGR J17091-3624, 
where the wind velocity was estimated as 0.03 c \citep{king2012}.      
Additionally, photoionisation modeling can be done using 
grid of models computed by {\sc xstar}  \citep{kallman2001}
if the quality of spectrum is good enough. Nevertheless, this is 
always done after continuum fitting is established,
and usually at least two or more photoionisated slab components are
needed to model all lines \citep{king2012}. In case of black hole 
transient IGR J17091-3624,
EWs of Fe{\sc xxv} and Fe{\sc xxvi} lines are 
approximately twice bigger when fitting {\sc xstar} models, 
than simple the Gaussian shape.   

In this paper, we present alternative analysis  of {\it Suzaku} XIS
X-ray spectra of \KU\, presented in \citet{kubota2007}.
Instead of multi-blackbody disk emission {\sc diskbb} with two Gaussians
fittet separately, we used
our atmospheric disk emission models ({\sc atm}) recently computed and 
published in \citet{rozanska2011}. Our theoretical spectra were obtained
from careful radiative transfer computations
in accretion disk atmospheres. All calculations of continuum and line 
spectra were performed simultaneously, taking into account relativistic
Compton scattering of X-ray photons. Absorption profiles of
Fe{\sc xxv} and Fe{\sc xxvi} spectral lines were carefully computed
as a convolution of natural, thermal (Gaussian) and pressure broadenings.
Expressions describing pressure broadening of iron lines were 
deduced from considerations by \citet{griem1974}.
In the case of our models lines are fitted together with underlying 
continuum.

The paper is organized as follows: Sec.~\ref{sec:src} describes recent 
observational status of Fe absorption lines in X-ray binaries, 
Sec.~\ref{sec:theor} explain principal assumptions of our theoretical 
model. All the procedure of data reduction and spectral fitting which
were done by us is presented in Sec.~\ref{sec:obs}. Conclusions and discussion 
are formulated in Sec.~\ref{sec:summary}.

\section{Narrow Fe absorption lines in X-ray binaries}
\label{sec:src}

Different types of X-ray binaries exhibit presence of narrow 
absorption lines from highly ionized iron. 
Tab.~\ref{tab:ratio} is an updated version of the Table 5 from 
\citet{boirin2004}. In the past research most of objects were fitted 
using {\sc powerlaw} 
model to describe continuum, and Gaussian shape for lines.
In some sources, being in soft state (i.e. dominated by the disk 
component) multi-temperature disk black body model ({\sc diskbb}),
and Voigt profiles for individual lines ({\sc kabs})  
were considered \citep{kubota2007}.     

Gaussian profile fitted to an individual absorption line puts constrains 
on the ionic column density. Such ionic column 
densities can be incorporated with photoionisation 
calculations to derive ionization parameter and total column density 
of absorbing material. In case of X-ray binaries,  
photoionisation models show $N_H$ of the order of $10^{23-24}$ cm$^{-2}$ 
with the value of ionization parameter $\xi = 10^{3-4}$. Such estimates
were obtained for a few systems, assumed to be highly inclined 

Usually prominent K$_\alpha$ absorption lines from He and H-like iron
are detected with EWs from -5 up to -61 eVs (minus sign denotes 
an absorption line). Only in few systems K$_\beta$ lines are visible.  
The extreme case is XB1916-053, in which He-like iron line changes its 
EW during the persistent and dipping time intervals. In persistent phase,
$EW=-30^{+8}_{-12}$, while during dipping phase the 
equivalent width  changes from $EW=-74^{+20}_{-27}$ to $EW=-168^{+44}_{-46}$ 
\citep{boirin2004}.
Summary of the objects with absorption lines and the connection of 
EWs with spectral state is shown in \citet{ponti2012}.  

In GX 13+1 properties of the absorber do not vary strongly 
with the orbital phase, which suggests that ionized plasma has a 
cylindrical geometry . Therefore, inclination of  GX 13+1 
was constrained  between 60-80$^{\circ}$.
Note, that almost all values of inclinations in fourth column of 
Tab.~\ref{tab:ratio} were not known before the discovery of iron absorption 
lines in those sources. 
Adopting wind geometry, the presence of Fe absorption lines
suggests high  inclination of those sources.
This is the reason why inclination numbers in fourth column are so similar.

\begin{table*}
\begin{center}
\caption{Measured absorption line properties and orbital parameters of sources 
with observed Fe absorption lines, L - LMXB, 
microq - micro-quasars, D - deeps, E - eclipses,
B - bursts, qB - quasi-periodic bursts, T - transient, O - outburst, 
J - radio jet.}
\label{tab:ratio}
\begin{tabular}{lllllllll} 
\hline \hline 
Source  &  ref. & Type & $P_{orb} (h)$ & i ($^{\circ}$) & Fe{\sc xxv} 
K$\alpha$ & Fe{\sc xxvi} K$\alpha$ &  Fe{\sc xxv} K$\beta$  
&  Fe{\sc xxvi} K$\beta$\\
& & & & & E (keV) & E (keV) & E (keV) & E (keV) \\
& & & & & EW (eV) & EW (eV) & EW (eV) & EW (eV) \\
\hline \hline

& & & & & & & & \\

XB 1916-053 (NS)  & [1] & LMXB & 0.8 & $60-80$ & 
$6.65^{+0.05}_{-0.02}$ &  $6.95^{+0.05}_{-0.04}$  & 
$7.82^{+0.03}_{-0.07}$ (a)  &  $8.29^{+0.08}_{-0.12}$ \\
 & &D, B & & &  $ -30^{+8}_{-12}$  & $-30^{+11}_{-12}$ &
$-21^{+9}_{-12}$   & $-19^{+11}_{-12}$ eV \\

 & & & & & & & & \\

GX 13+1 (NS)  & [2] &  LMXB & 577.4  & $ 60-80 $ (b)  &
$6.75^{+0.02}_{-0.05}$  & $7.02 \pm 0.06$ &  $7.91^{+0.46}_{-0.15}$ 
  &  $8.24^{+0.04}_{-0.06}$ \\ 
& & Atol, B&  & & $ -20 \pm 5$ & $-41 \pm 7 $ & $-33 \pm 12 $ & $-37 \pm 15 $ \\

& & & & & & & & \\

MXB 1659-298 (NS) & [3] & LMBX& 7.1 & $\sim 80$ & 
$6.64 \pm 0.02 $ & $6.90^{+0.02}_{-0.01}$ & -- & -- \\
& &D, E, T & & &  $ -33^{+9}_{-20}$  & $-42^{+8}_{-13}$ & --  & -- \\

& & & & & & & & \\ 

X 1254-690 (NS)  & [4] & LMXB  & 3.9 & $ 60-80$  & -- &
$6.95 \pm 0.03 $  & -- & $8.20^{+0.05}_{-0.10}$ \\
& &D & & &  --  & $ -27^{+11}_{-8}$  & --  & $ -17 \pm 9$  \\

& & & & & & & & \\

X 1624-490 & [5] & LMXB & 20.9 & $60-80$ & $6.72 \pm 0.03 $ &  
$7.0 \pm 0.02 $ & -- & -- \\
& &D & & & $ -7.5^{+1.7}_{-6.3}$  & $ -16.6^{+1.9}_{-5.9}$  
& --  & --  \\

& & & & & & & & \\

XB 1323-619 & [6] & LMXB & 2.93  & $60-80$  &  $6.70 \pm 0.03 $ 
& $6.98 \pm 0.04 $  & -- & -- \\
& &D, qB & & & $ -22 \pm 3 $  & $ -27 \pm 4 $  & -- &  --  \\

& & & & & & & & \\

Cir X-1 (NS) & [7] & B, D, O, J & 398 & high & $6.701 \pm 0.009 $  &
 $6.965 \pm 0.01$   & -- &-- \\ 

& & & & & & & & \\

4U 1630-472 (BH) & [8] & LMXB & 690 & $\sim 70$ & 
$6.72 \pm 0.01 $  & $6.980^{+0.009}_{-0.007}$ &   
$7.84^{+0.04}_{-0.03}$  & $8.11^{+0.05}_{-0.06}$ \\
& &D, T, O & & & $ -21 \pm 3$  & $ -30^{+2}_{-3}$  
& $ -18.2^{+4}_{-6}$   &  $ -15.8^{+5}_{-7}$  \\

& & & & & & & & \\

GRO J1655-40 (BH) & [9] &  microq  & 62.9 (c) & $ 69-85 $ & 
$6.63 \pm 0.07 $ &  $6.95 \pm 0.10$  & $ 7.66 \pm 0.13 $ & -- \\
& &D, O & & & $ -61^{+15}_{-13}$  & $ -25^{+13}_{-11}$  
& $ -35^{+30}_{-29}$  & --  \\

& & & & & & & & \\

 & [10] &   &  &  & $6.698 \pm 0.001 $ &  $6.999 \pm 0.002$  & 
$ 7.842 \pm 0.005 $ & $ 8.277 \pm 0.01 $ \\
& & & & & $ -45.59 \pm 1.8$  & $ -30.82 \pm 2.37$  
& $ -48.11 \pm 4.46 $  & $-27.63 \pm 5.53$  \\

& & & & & & & & \\

GRS 1915+105 (BH) & [11] & microq & 804 & $\sim 70$ & 
$6.72^{+0.015}_{-0.017}$  & $7.000^{+0.020}_{-0.014}$ &   
$7.843^{+0.037}_{-0.035}$  & $8.199^{+0.065}_{-0.064}$ \\
& & & & & $ -35.8^{+3.7}_{-7.8}$  & $ -43.2^{+6.3}_{-5.3}$  
& $ -30.8^{+11.2}_{-5.8}$   &  $ -22.^{+11}_{-12}$  \\

& & & & & & & & \\

 & [12] & & & & $6.684 \pm 0.022$  & $6.984 \pm 0.009 $  & --  
& -- \\
& & & & & $ -6$  & $ -18$  & --  &  --  \\

& & & & & & & & \\

H1743-322 (BH) & [13] & microq & -- & $\sim 70$ & 
$6.709 \pm 0.004 $  & $6.981 \pm 0.004$ &  -- & -- \\
& &D, T, O & & & $ -5 \pm 1$  & $ -20 \pm 2 $  & --  & -- \\

& & & & & & & & \\

IGR J17091-3624 (BH) & [14] & T, O & -- & -- & 
$6.91 \pm 0.01 $  & $7.32^{+0.01}_{-0.06}$ &  -- & -- \\
& & & & & $ -21^{+0.5}_{-0.2}$  & $ -32^{+1.8}_{-0.4} $  & --  & -- \\

& & & & & & & & \\

\hline \hline
\end{tabular}
\end{center}

[1] -  XMM, \citet{boirin2004}, [2] - XMM, \citet{ueda2001,trigo12},
[3] - XMM, \citet{sidoli2001}, [4] - XMM, \citet{boirin2003},
[5] - XMM, \citet{parmar2002}, [6] - XMM, \citet{church05}, 
[7] - {\it Chandra}, \citet{brandt2000}, [8] - SUZAKU epoch 4 
\citet{kubota2007}, [9] - ASCA, \citet{ueda1998}
[10] - {\it Chandra}, \citet{miller2008},
 [11] - ASCA, \citet{kotani2000},
[12] - {\it Chandra}, \citet{lee2002}, [13] - {\it Chandra} epoch 1,
\citet{miller2006}, [14] - {\it Chandra}, \citet{king2012},

(a) - Identified as S{\sc xvi} and Ni{\sc xxvii} transitions by 
       \citet{boirin2004}

(b) - Low $i$ suggested by high radio to X-ray ratio \citep{schnerr03},
    but high $i \simeq 60-80^{\circ}$ by detection of Fe abs. lines
    \citep{trigo12}

(c) - From optical observations \citep{orosz1997}

\end{table*}

Properties of the absorption features in GX 13+1, and  MXB 1659-298
show no obvious dependence on the orbital phase, except during a dip from 
X1624-490, where there is an evidence for the presence of additional colder
material. 
\citet{boirin2003} have modeled narrow absorption features of 
Fe{\sc xxvi}  $K_{\alpha}$ and $K_{\beta}$ during non-dip intervals.
Therefore, ionized absorption features may be common characteristics 
of disk accreting systems. Systems with shorter orbital periods are
expected to have smaller accretion disks. 

In XB 1323-619 a number of absorption lines (6.7 and 6.9 keV) were discovered 
in non-dip, non-burst spectra \citep{church05}. Curve of 
growth analysis provided a consistent solution in which line ratio
was reproduced assuming collisional ionization with kT=31 keV, close 
to the accretion disk corona (ADC) electron temperature in this source. 
Thus, \citet{church05} proposed that those absorption lines in the 
dipping low mass X-ray binaries (LMXB) are produced in ADC.

From observational point of view ADC can produce the same 
spectral features like upper layers of 
accretion disk atmosphere. Temperature of both emission regions can be
comparable as we explain in the following Sections. 

\section{Fe absorption lines in accretion disk atmosphere}
\label{sec:theor}

In this paper, we fitted the few iron rich model disk atmospheres, which were 
already computed and presented in \citet{rozanska2011}
to the {\it Suzaku} X-ray spectra of \KU\ .  Since the observed iron lines
in \KU\ were discovered in absorption, we use only non-irradiated 
modeled disk spectra. 

Full simulation of emission from accretion disk atmospheres is very time
consuming and depends on the assumed global disk model.
The code used in this paper 
was fully described in our previous paper \citet{rozanska2011}.
We remind here the most essential properties of our numerical approach. 
The accretion disk at a given accretion rate and black hole spin 
is divided on approximately 15 rings. At each ring we computed local 
model atmosphere using complete and consistent radiative transfer code
ATM21 \citep{disk2008}. The code assumes that the vertical optical depth 
at some standard photon frequency is the independent variable in the model.
At the fixed chemical composition model atmosphere assumed two parameters:
the effective temperature, which at each ring was determined from 
the accretion rate, and the gravity in vertical direction caused by the 
black hole for the given mass. Each model atmosphere was strongly non-gray, 
and the local emissivity or opacity was a sum of non-gray bound-free, 
free-free, and bound-bound absorption/emission caused by individual 
ions. Moreover, local opacity and emissivity was influenced by 
Compton scattering by free electrons. We always assumed that a single 
local electron temperature is valid for all processes, also those including 
ions. Velocity distribution of all particles is relativistic and Maxwellian.

We numerically simulated emission assuming radial effective 
temperature distribution for slim disk model \citep{abramowicz88}  
as the most reasonable one. Nevertheless, calculations of $T_{eff}(R)$
for the standard SS disk \citep{SS73} or for NT disk \citep{novikov73} give 
similar results, as was shown in Fig. 1 by \citet{rozanska2011}. 
In all cases, for the mass of black hole
$M_{BH}=10 M_{\odot}$ and for high accretion rate, temperature 
in the inner disk can reach $10^7$ K. This temperature even rises
when spin of the black hole increases.  

The whole disk was divided into individual rings and at each distance
from the black hole  full radiative transfer computations were done 
assuming that the atmosphere is in radiative and hydrostatic equilibrium. 
We use our code for model atmosphere computations, ATM21, suitable for 
computing disk vertical structure and its X-ray intensity spectra. 
Compton scattering on free electrons and ionization structure 
were self consistently taken into account \citep{madej2004}. Final 
spectrum of the disk was obtained by integrating of contributions from
individual rings. 

In particular our X-ray model spectra included profiles of eight 
most prominent lines from highly ionized iron. Those lines belong to two 
fundamental series of iron.  Helium-like iron Fe{\sc xxv} produces 
the resonant line at E=6690 eV,
commonly denoted as 6.7 keV, and three lines at energies:
E=7881, 8226, and  8416 eV. The series of hydrogen-like iron 
Fe{\sc xxvi} consists of the resonant line at E=6898 eV and higher lines
at E=8172, 8621, and 8829 eV.

\begin{table*}
\begin{center}
\caption{Equivalent widths of the model iron line components (in eV):
FeXXV and FeXXVI resonant lines for five sets of the couple of 
parameters $\dot m$ and spin $a$. All EWs are given  for 
viewing angles from $i=11.4^{\circ}$ (face-on disk),
to $i=88.9^{\circ}$ (edge-on disk). In the first row we put global parameters
of a model: accretion rate $\dot m$ in units of Eddington accretion rate, 
and dimensionless black hole spin $a$.
}
\label{tab:eqw}
\begin{tabular}{|c|cc|cc|cc|cc|cc|}
\hline \hline
  & $\dot m$=0.01 &  a=0.98  & $\dot m$=0.1 &  a=0.98  & $\dot m$=0.01 &  a=0  &
 $\dot m$=0.1 & a=0  & $\dot m$=1 &  a=0  \\
 inc.($^{\circ}$)  & FeXXV & FeXXVI  & FeXXV & FeXXVI  & FeXXV & FeXXVI  & FeXXV &
FeXXVI  & FeXXV & FeXXVI  \\
\hline
11.4  & -20.5 & -10.5  & -7.5 & -3.6  & -56.7 & -1.7  & -50.0 & -1.7  & -41.1
& -9.7  \\
26.1  & -20.1 & -10.3  & -7.4 & -3.5  & -54.7 & -1.6  & -49.4 & -1.8  & -40.8
& -9.5  \\
40.3  & -19.4 & -9.9  & -7.0 & -3.1  & -50.4 & -1.4  & -48.4 & -1.8  & -40.0 &
-9.1  \\
53.7  & -18.2 & -9.3  & -6.6 & -2.7  & -44.0 & -1.1  & -46.8 & -1.9  & -38.5 &
-8.8  \\
65.9  & -16.4 & -8.4  & -5.9 & -2.1  & -37.9 & -0.9  & -44.6 & -2.0  & -35.9 &
-8.6  \\
76.3  & -13.8 & -7.5  & -4.9 & -1.6  & -34.7 & -0.8  & -41.6 & -2.2  & -31.7 &
-8.9  \\
84.2  & -10.4 & -6.3  & -3.7 & -1.2  & -34.5 & -0.7  & -36.9 & -2.6  & -25.2 &
-9.8  \\
88.9  & -6.8 & -4.7  & -2.4 & -1.1  & -36.5 & -0.6  & -26.9 & -4.3  & -15.4 &
-10.6  \\
\hline \hline
\end{tabular}
\end{center}
\end{table*}

Additionally, we include three bound-free edges: Fe{\sc xxiv} 
edge at E = 8690 eV,
Fe{\sc xxv}  edge at E= 8829 eV, and Fe{\sc xxvi} edge at E=9278 eV, which are
visible in high energy X-ray band. These are ionization energies 
from the innermost K-shell. Other iron edges, which were included in our
computations arise from all K, L, M shells, but we do not list them since they
are not prominent above 2 keV. 

\citet{rozanska2011} have shown that in galactic black hole binaries, 
accretion disk must be very hot, with local effective temperatures 
approaching $10^7$ K on the innermost orbits. In such high temperatures 
iron is almost fully ionized and spectral features of helium and
hydrogen-like resonant iron lines must play a crucial role in interpreting 
of the observed GBHB spectra in soft state. 

Therefore, iron absorption lines, which were observed recently in disk 
dominated binaries, see section above, most likely are resonant lines from
upper layers of accretion disk atmospheres. In our models we take into 
account all three line broadening mechanisms: natural, thermal (Gaussian)
and pressure broadening. Broadened iron line profiles were carefully computed
as the convolution of all three partial broadenings.  

Tab.~\ref{tab:eqw} presents theoretical equivalent widths of 
resonant lines computed
in our radiative transfer simulations \citep{rozanska2011}. All were 
computed only for five sets of global disk parameters and for eight 
viewing angles. In all cases mass of the black hole was assumed to be
equal 10$M_{\odot}$, nevertheless we do not have mass estimation in \KU\. 
Models do not cover all values in the parameter space, but 
in this paper for the first time we present fitting to the \KU\ 
absorption lines together with underlying continuum.


Note, that theoretical EWs are usually not very high, and only for the 
hottest model  the ratio of EWs for Fe{\sc xxv} and Fe{\sc xxvi} lines
is of the order of 2 (four leftmost columns in Tab.~\ref{tab:eqw}). 
For other models the atmospheric temperature is too low to produce 
prominent hydrogen-like iron line. In future work we plan to compute the whole
grid of disk atmosphere models and make them suitable for spectral fitting. 

\section{Suzaku observations of \KU\ }
\label{sec:obs}

\subsection{The source}

\begin{figure}[t]
\includegraphics[width=70mm,angle=-90]{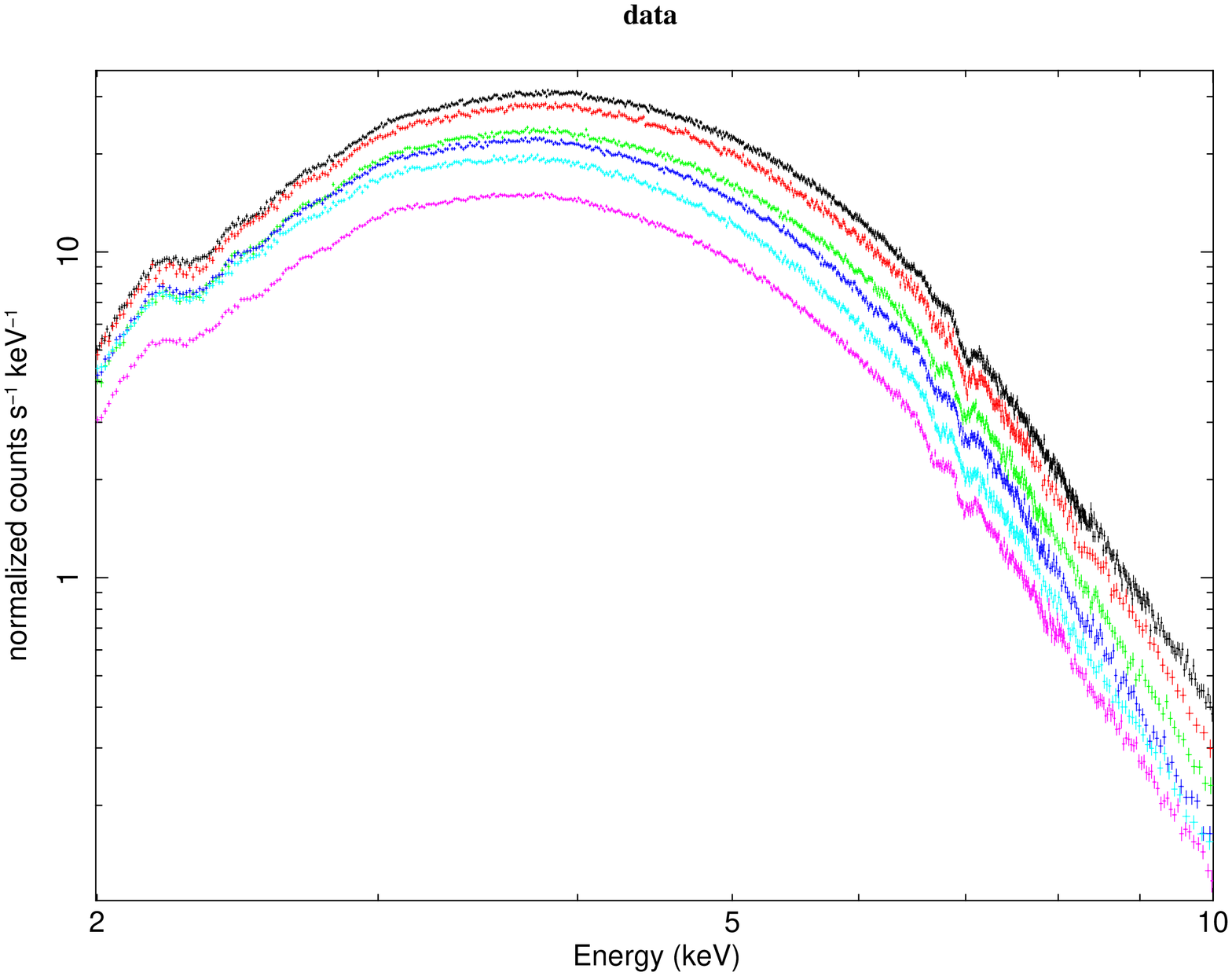}
\includegraphics[width=85mm]{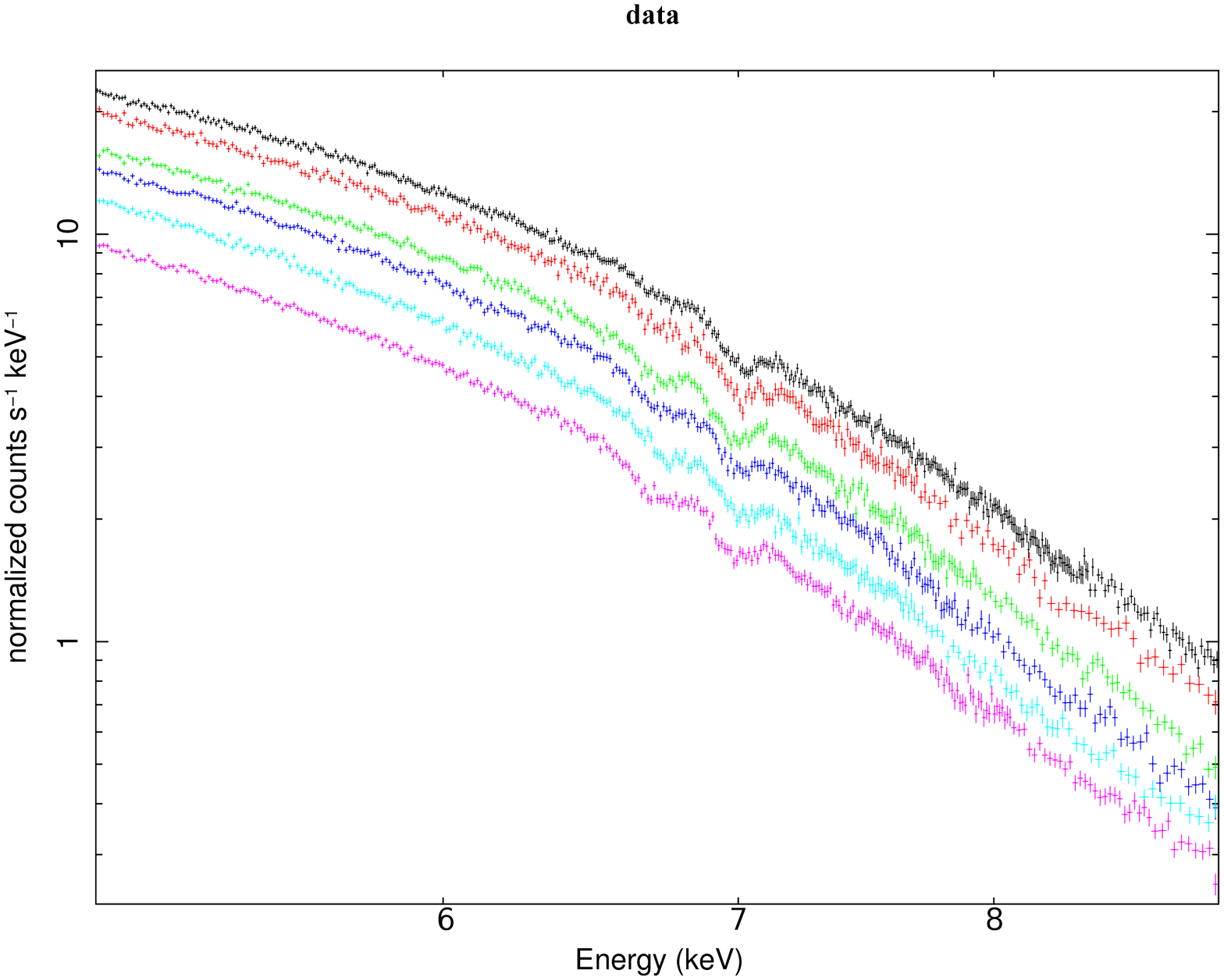}
\caption{X-ray spectra of \KU\ obtained with the XIS023 showing prominent 
bump  similar to multi-temperature disk emission - upper panel. 
Lower panel is a focus on the iron line region from 5-9 keV, where absorption 
features are clearly seen.Colors mean different epochs 
of observations starting from black - epoch 1, up to magenta - epoch 6.
The data are very similar to those presented
in \citet{kubota2007} in Fig.2 and Fig.5.}
\label{fig:data}
\end{figure}

Black hole candidate \KU\ is known for X-ray outbursts that repeat within 
roughly 600-690 days \citep{jones76,parmar95}. The source has been 
observed by many major X-ray missions, showing heavy absorption with 
$N_H= 5-12 \times 10^{22} $ cm$^{-2}$ \citep{tomsick98}. X-ray absorption dips
observed in this source suggest large inclination angle around $60^{\circ}$
\citep{kuulkers98}. This conclusion was based on the model where dips are 
produced during eclipse of accretion disk corona (ADC) by warm bulge, which 
developes when matter falls from companion star on the outer 
rim of the disk. 

No optical counterpart is known for \KU\ mostly due to its large reddening 
and location in a crowded star field \citep{parmar86}. Therefore, 
the distance and mass of the compact object in \KU\ are unknown and the 
source is classified as a black hole, due to the similarity of its X-ray 
spectral and timing properties to those shown by systems with measured black hole 
masses. 

Recent {\it Suzaku} monitoring of the source led 
to successful detection
of absorption lines in all six observations \citep{kubota2007}.
Lines were interpreted as an absorption in the wind. In this paper 
we show that this is not the only one interpretation. 
By fitting iron absorption lines produced in outer regios of the accretion disk
and hot atmosphere, together with 
underlying continuum, we demonstrate that existing satellites are not able to 
distinguish between both models. The model of outflowing wind with velocity 
of the order of hundred km/s fits as well as the model of static atmospheres. 
Our analyze also suggests that the wind can originate from hot accretion disk 
atmosphere.

\subsection{Data reduction}
\label{sec:suz}

We have downloaded and reduced the same {\it Suzaku} data of \KU\ as those 
published by \citet{kubota2007}. Six observations were performed in time 
period from 08.02.2006 to 23.03.2006 after an outbursts that was reported 
in 12.2005 by {\it RXTE} ASM. The exact time of 
observations made by {\it Suzaku} in 
comparison with ASM lightcurve is marked
in \citet{kubota2007} Fig.1 by vertical arrows.  

The source was observed by X-ray Imaging Spectrometer(XIS) 
operating in the  0.2 - 12 keV energy range, and one of two 
Hard X-ray Detectors (HXD), both instruments on the board of {\it Suzaku}. 
Nevertheless, for this analysis we took
only XIS data. XIS is build out of four CCDs, located in the focal
plane of X-ray mirrors (XRT). 
Three of those CCDs named: XIS0, XIS2, and XIS3 are front illuminated (FI), 
while XIS1 sensor is back illuminated (BI). 
In this paper we analyze both FI and BI CCD data. 
Due to high brightness of \KU\ , XIS was set to 1/4 window option 
(which means we have 2 sec read out time) and first 
four observations were performed in burst option. $2 \times 2$ and 
$3 \times 3$ editing modes 
were used in observations.
All exposure times and count rates are presented in Tab.~\ref{tab:obser}.

In further analysis we used only $3 \times 3$ editing mode data 
because $2 \times 2$ data 
were not well calibrated during observation time. 
Although background was negligible during all 6 observations we subtracted 
it from the data. 
For extracting spectra  from {\it Suzaku} filtered event file we used  
{\it xselect} - a HEAsoft tool. 
Before extracting spectra or light curves in case of a point source we have 
to create circular extraction region centered on the sources. 
It is recommended to create regions which encircles 99 \% of point 
source flux. In later reduction process we used {\it xissimarfgen} script 
to build response files for our spectra. 
{\it Xissimarfgen} is a generator of auxiliary response files (ARFs) that 
is based on ray-tracing and is designed for the analysis data 
from {\it Suzaku} XIS detectors.
 This script is calculating ARFs through Monte-Carlo simulations. 
The idea is that program ray-traces X-ray photons through the XRT and XIS and 
counts the number of events detected in extraction regions
defined by user. 
We need to assume a sufficient number of photons to limit statistical 
errors to good level. In our analysis we assumed number of simulated 
photons to 400000 to minimize Poisson noise in ARFs.  In order to carry 
out further analysis we downloaded redistribution matrix files (RMFs) from
CALDB catalogs (for appropriate dates).

As we have created ARFs and collected all RMFs files, we used an ASCA tool 
called {\it addascaspec} to combine source and background spectra 
and responses of {\it Suzaku} XIS0, XIS2, and XIS3 (FI chip) data. 
We also have tried to add all FI and BI (XIS1) chips despite that such a 
procedure was not recommended but we thought it can increase statistics 
in softer part of spectra.

We found out that over all six observations count rate declined 
through the time. We know that there is a count rate threshold above which 
data from the source can suffer from photon pileup effect. 
In case of  point source, it is $\sim 100$ counts/sq arcmin/s (CCD exposure) 
for  {\it Suzaku} XIS instrument. 
In all six observations count rate for \KU\ was 
around 100 counts/sec., so we can assume both:  that data have suffered 
from photon pileup or not. 

To minimize pileup effect we have used two externally contributed routines 
{\it aeattcor.sl} and {\it pile\_estimate.sl} 
\footnote{http://space.mit.edu/ASC/software/suzaku/}. 
Both scripts are designed for the analysis 
of {\it Suzaku} data that might be affected by photon pile-up effect. 
Script {\it aeattcor.sl} 
improves the attitude correction for the slow wobbling of the optical
axis for bright sources by using their detected image to create a new 
attitude file.  Next, script {\it pile\_estimate.sl} 
is designed to be run just after {\it aeattcor.sl}. 
It creates an image that will show an estimated, minimum pile-up fraction 
for user specified levels. This allows us to create a region file that
excludes the most piled areas, and then estimate the effective pile-up 
fraction of the remaining events. 

But data with removed pile-up effect had much poorer statistics than those
without removal, so in further modeling we have used not cleared data. 
\KU\ is near the threshold for pile-up, but we also can assume that there 
is no pile-up in those data as it was done by \citet{kubota2007}. 
All other data reduction and calibration uncertainties were done in the same
way as in \citet{kubota2007} paper. Extracted XIS spectra for six
observations and the region of iron absorption lines are 
presented in Fig.~\ref{fig:data}. Final exposure time
and count rate in each XIS CCD chip for each epoch of observation is 
presented in Tab.~\ref{tab:obser}.

\begin{table}[tp]
\begin{center}
\caption{{\it Suzaku} observations of \KU\ for all  6 Epochs of the year 2006. 
Dates of observations are listed in the first column }
\label{tab:obser}
\begin{tabular}{lllll} 
\hline \hline
Epoch   & XIS  & $T_{exp}$ [ks] & count rate \\ 
\hline
   & 0   & 11.08 & 8.87E+01 $\pm$ 8.95E-02 \\ 
1  & 1  & 11.06 & 8.79E+01 $\pm$ 8.92E-02 \\
Feb 8   & 2   & 11.02 & 1.09E+02 $\pm$ 9.94E-02 \\
   & 3   & 11.08 & 1.28E+02 $\pm$ 1.08E-01 \\
\hline 
\hline
  & FI    & 33.17 & 1.02E+02 $\pm$ 1.10E-01 \\
  & FI+BI & 44.23 & 9.86E+01 $\pm$ 7.60E-02 \\

\hline \hline
   & 0   & 4.79 & 7.48E+01 $\pm$ 1.25E-01 \\ 
2  & 1   & 4.86 & 7.37E+01 $\pm$ 1.23E-01 \\
Feb 15  & 2   & 4.76 & 9.06E+01 $\pm$ 1.38E-01 \\
   & 3   & 4.79 & 8.95E+01 $\pm$ 1.37E-01 \\
\hline 
\hline  
  & FI    & 14.34 & 8.41E+01 $\pm$ 8.09E-02 \\
  & FI+BI & 19.20 & 8.13E+01 $\pm$ 6.84E-02 \\

\hline \hline
   & 0   & 10.7 & 5.73E+01 $\pm$ 7.32E-02 \\ 
3  & 1  & 10.71 & 5.84E+01 $\pm$ 7.38E-02 \\
Feb 28   & 2   & 10.64 & 7.40E+01 $\pm$ 8.34E-02 \\
   & 3   & 10.7 & 6.71E+01 $\pm$ 7.92E-02 \\
\hline 
\hline 
  & FI    & 32.04 & 6.51E+01 $\pm$ 4.71E-02 \\
  & FI+BI & 42.75 & 6.36E+01 $\pm$ 4.00E-02 \\
\hline 
\hline 
   & 0   & 10.3 & 5.87E+01 $\pm$ 7.55E-02 \\ 
4  & 1  & 10.28 & 5.30E+01 $\pm$ 7.18E-02 \\
Mar 8   & 2   & 10.25 & 6.54E+02 $\pm$ 7.99E-02 \\
   & 3   & 10.3 & 6.44E+02 $\pm$ 7.91E-01 \\
\hline 
\hline 
  & FI    & 30.85 & 6.24E+01 $\pm$ 4.68E-02 \\
  & FI+BI & 41.13 & 5.99E+01 $\pm$ 3.96E-02 \\
\hline 
\hline 
   & 0   & 5.24 & 5.29E+01 $\pm$ 1.01E-01 \\ 
5  & 1  & 23.18 & 5.35E+01 $\pm$ 4.80E-02 \\
Mar 15   & 2   & 5.24 & 6.02E+01 $\pm$ 1.07E-01 \\
   & 3   & 5.24 & 5.82E+01 $\pm$ 1.05E-01 \\
\hline 
\hline 
  & FI    & 15.73 & 5.65E+01 $\pm$ 6.39E-02 \\
  & FI+BI & 38.9 & 5.43E+01 $\pm$ 3.91E-02 \\
\hline 
\hline
   & 0   & 15.37 & 4.45E+01 $\pm$ 5.38E-02 \\ 
6  & 1  & 21.65 & 4.64E+01 $\pm$ 4.63E-02 \\
Mar 23   & 2   & 15.36 & 5.33E+01 $\pm$ 5.89E-02 \\
   & 3   & 15.36 & 5.04E+01 $\pm$ 5.73E-02 \\
\hline 
\hline
  & FI    & 52.37 & 4.91E+01 $\pm$ 3.40E-02  \\
  & FI+BI & 67.74 & 4.81E+01 $\pm$ 2.76E-02 \\
\hline 
\hline 
\end{tabular}
\end{center}
\end{table}

\subsection{Spectral fitting}

Our numerical disk spectra \citep{rozanska2011} were converted to 
the FITS format, suitable for XSPEC fitting package (version 12.6). 
Atmospheric models, hereafter named {\sc atm} were rare in 
parameter space. 
Since we do not know $M_{BH}$, $\dot m$ and $a$ of the source, 
we took a prior set of parameters $\dot m$ and $a$ for a range of 
inclinations and fixed mass $M_{BH}=10 M_{\odot}$. 
We  did not aim here  to derive  neither accretion rate nor spin of the 
black hole. We for the first time fitted  iron absorption 
line profiles different than simple Gaussian.

It was almost obvious that the best model is the 
hottest one, obtained for $\dot m=0.01$ in Edd. units and spin $a=0.98$,
since EWs of iron line were the biggest (see two leftmost 
columns in Tab.~\ref{tab:eqw}). After that, we were able to fit observations 
taking viewing angle as the only free parameter.   

\subsubsection{Continuum}
\label{sec:cont}

First, we performed  fitting over the whole energy range 2-9 keV. 
Instead of using multi-temperature XSPEC model {\sc diskbb},
we applied our atmospheric disk emission {\sc atm} multiplied by 
interstellar absorption {\sc wabs} model.  
Additionally we have added {\sc powerlaw}  model 
to mimic possible influence of irradiation at hard part of the spectrum. 
From theoretical point of view, in sources being in soft 
state, dominated by disk emission,  we expect that irradiation only 
slightly change hard tail of the disk spectrum. From instrumental 
point of view, high energy part can be affected by pile-up 
(see Sec.~\ref{sec:suz}). In \citet{kubota2007} this effect was 
modeled by a broad Gaussian in emission at 10 keV. 
In our fitting procedure, the {\sc powerlaw} model with flat photon 
index should balance those two effects. Following suggestions made 
by \citet{kubota2007} we have added Gaussian at $\sim 3.2$ keV with 
width of the order of 0.1 keV to reduce systematic errors arising from 
current uncertainties in the instrument response at the gold M-{\sc ii} edge. 

\begin{figure}[t]
\includegraphics[width=85mm,angle=0]{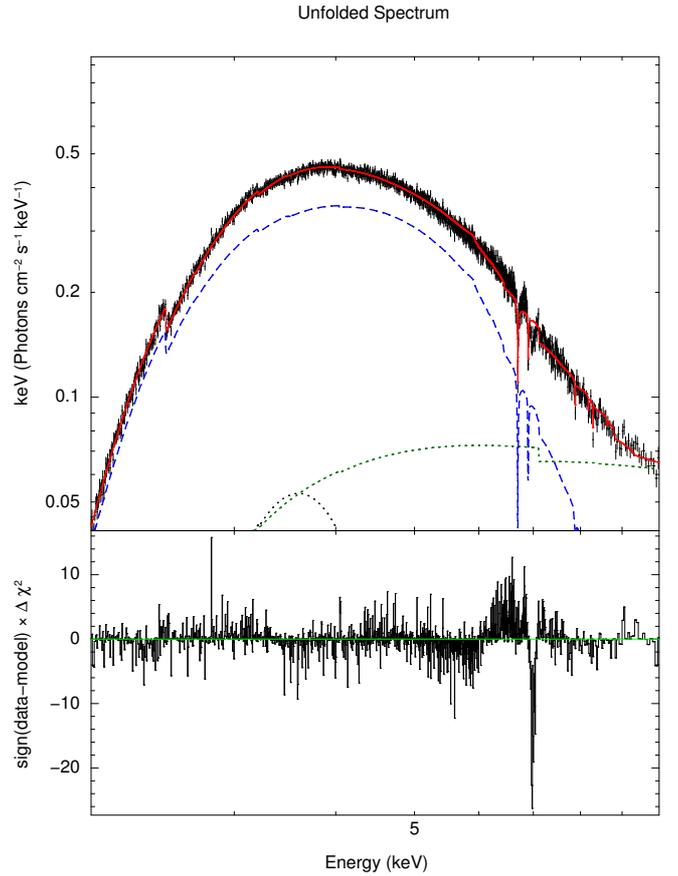} 
\caption{Fitting of atmospheric disk model to the epoch 3 data in the
range 2-9 keV. Observed data, black crosses, are a sum of three front 
illuminated (XIS0,2,3) and one back illuminated (XIS1) CCDs. The total model
{\sc wabs}*({\sc atm+pl + gaussian}) is represented by red line. Separate
model components are represented by long-dashed blue line - {\sc atm}, 
short-dashed green line - {\sc pl}, and dotted black line - instrumental 
{\sc gaussian}. 
Reduced  $\chi ^2/d.o.f. =  2618.57 / 2182 = 1.2$ for the total model.}
\label{fig:cont}
\end{figure}
 
We point out here that our {\sc atm} atmospheric disk emission 
deviates from the multi-blackbody disk spectrum, since we took into 
account Compton scattering in hot disk atmosphere \citep{gancarczyk2011}.
We did not assume anything about the total disk hardening factor.
The effect of spectrum hardening is self consistently computed in our 
numerical model.  

There are two ways to obtain high effective temperature in the 
inner rings of the accretion disk. First, high $T_{\rm eff}$ is the result
of high accretion rate. 
This is quite obvious, higher accretion rate implies generation of higher
flux of radiation especially in inner rings. Therefore, the value of
accretion rate affects the most energetic part of the 
spectrum. 
The second way is to spin up black hole. If the black hole rotates than 
inner rings radiate more energy since the marginally stable orbit moves 
towards the black hole. Those rings are hot and, therefore, the total
continuum spectrum looks as if the disk 
accreted at a high accretion rate. Additionally, disk spectra have slightly 
different shape for various viewing angles. To distinguish what particular
spectrum really fits the data we need much more models. For the \KU\ 
observed data favour the hottest model. However, we cannot claim about 
the accretion rate and black hole spin in this source. This is rather 
because absorption lines in such a model are 
the strongest and agree with observations.

Taking into account that the source \KU\ is covered by interstellar 
gas of high column density of the order $8-9 \times 10^{22}$ cm$^{-2}$, 
and the data were slightly affected by pile-up, we present continuum 
fitting only for Epoch 3 data. Each other set of observations gave similar 
results.

Fig.~\ref{fig:cont} presents our continuum fitting for Epoch 3 of FI+BI 
coadded spectra. The absorbing column density from interstellar gas was 
fitted to $N_H = 8.23 \pm 0.08$, which is consistent with previous fits. 
The best fitted atmospheric model gives inclination $i= 11 \pm 5 ^{\circ}$ 
and normalization $1.50 \pm 0.08 \times 10^{-5}$. 
Additional {\sc powerlaw} photon index equals $\Gamma =  1.48 \pm 0.55$. 
In this fitting we didn't freeze any parameter. 
Our goodness of fit i.e. ${\chi}^2$ per degrees of freedom 
is $ 2618.57/ 2182 = 1.2 $.

\subsubsection{Iron line complex}

For further analysis we decided to proceed fitting in the energy band 
restricted to the region of iron line complex i.e. between
6 and 9 keV. In that narrow band we still fit lines together with 
underlying continuum as a single model, {\sc atm}. We do not make 
any additional assumption to the line or edge profile, 
since all features are already build up in our radiative transfer 
calculations of atmospheric disk emission.

\begin{figure*}
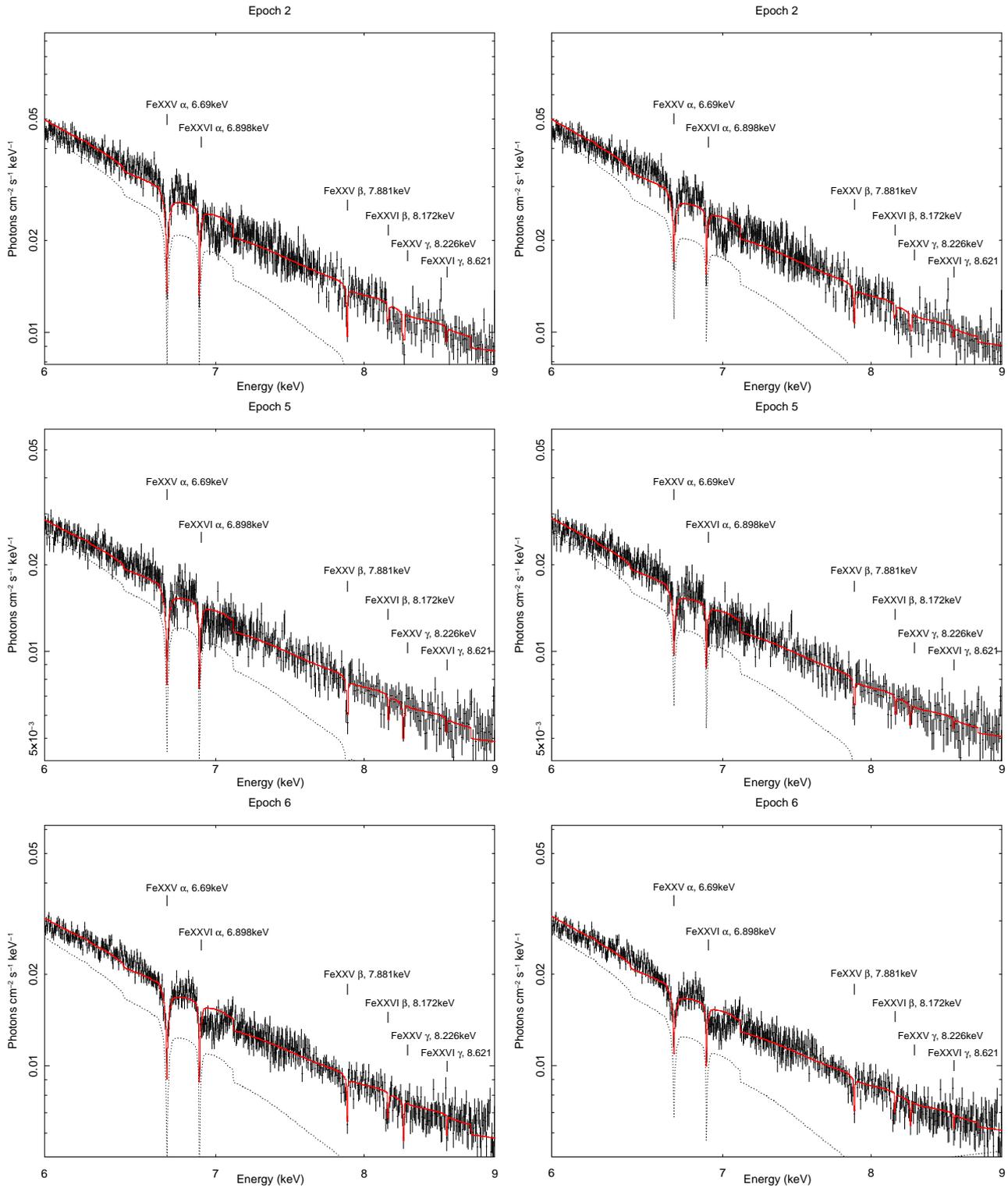

\begin{center}
\hbox{\includegraphics[width=67mm,angle=-90]{obs2_6_9_inc11_uf.ps}
\includegraphics[width=67mm,angle=-90]{obs2_6_9_inc70_uf.ps}}
\hbox{\includegraphics[width=67mm,angle=-90]{obs5_6_9_inc11_uf.ps}
\includegraphics[width=67mm,angle=-90]{obs5_6_9_inc70_uf.ps}}
\hbox{\includegraphics[width=67mm,angle=-90]{obs6_6_9_inc11_uf.ps}
\includegraphics[width=67mm,angle=-90]{obs6_6_9_inc70_uf_2.ps}}
\end{center}
\caption{Iron line region from 6-9 keV, where absorption 
features are clearly seen. Different panels represent observations
for three different epochs. 
Black crosses show the data,  black dotted lines represent 
{\sc atm} model, while red solid lines 
show total model. Upper panels show disk seen 
at $i=11\pm 5^{\circ}$, whereas lower panels show disk seen at 
$i=70 \pm 6^{\circ}$.
All spectra favor fitting for $i=11\pm 5^{\circ}$, since modeled
lines are the deepest for this model. }
\label{fig:band}
\end{figure*}

\begin{figure*}
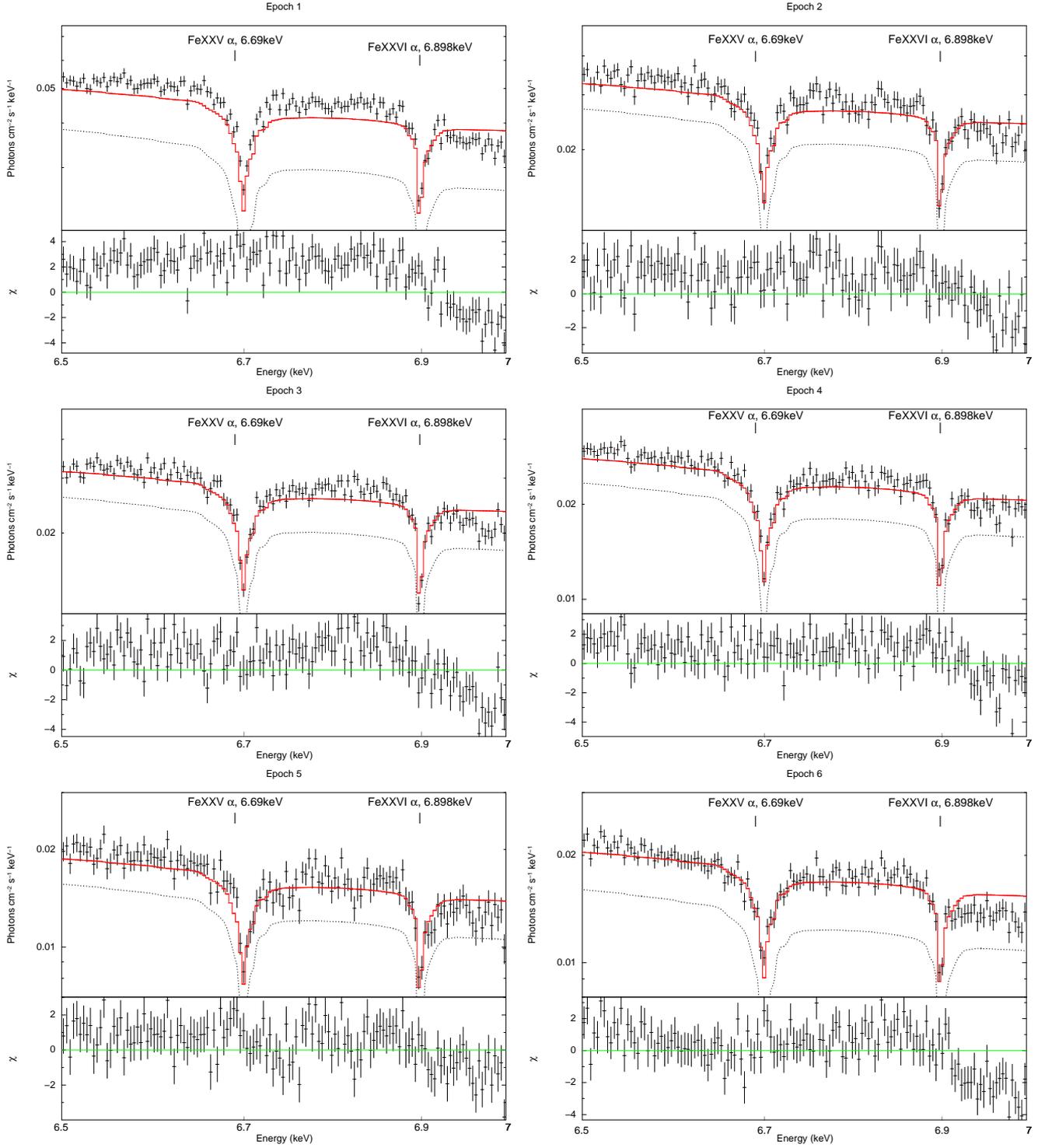

\hbox{\includegraphics[width=65mm,angle=-90]{linie1.ps}
\includegraphics[width=65mm,angle=-90]{linie2.ps}}
\hbox{\includegraphics[width=65mm,angle=-90]{linie3.ps}
\includegraphics[width=65mm,angle=-90]{linie4.ps}}
\hbox{\includegraphics[width=65mm,angle=-90]{linie5.ps}
\includegraphics[width=65mm,angle=-90]{linie6.ps}}
\caption{Focus on the resonant He and H-like iron lines for the fitting done 
in the 6-9 keV energy range (see text for explanation). 
Black crosses show the data, black dotted lines represent 
{\sc atm} model, while red solid lines 
show total model. Clearly, complicated 
line profiles computed in our model, match the observations very well for 
almost each epoch. Residues are presented below each panel. 
Here we present spectra for $i=11\pm 5^{\circ}$.}
\label{fig:line}
\end{figure*}   

\begin{table}[tp]
\begin{center}
\caption{Fitting results of our atmospheric disk emission model 
to the \KU\ {\it Suzaku} data  between 6-9 keV 
for the inclination angle $i=11 \pm 5^{\circ}$}
\label{fit:res}
\begin{tabular}{*{5}{l}}
\multicolumn{5}{ c }{Energy range 6-9 keV} \\
\hline \hline
 Ep. & WABS & ATM & PL & $\chi^2/d.o.f.$ \\
  & $N_H$ ($10^{22}$) & Inc. (ang) & Ph. Index &   \\
  &   & Norm ($10^{-5}$) & Norm ($10^{-3}$) &  \\
\hline
1 & 8 & $11 \pm 5$ & -0.7 & 2198.07/819 \\
  &   & 2.9 $\pm$ 0.6 & 3.46 $\pm$ 0.07 & = 2.68   \\
\hline
2 & 8 & $11 \pm 5$ & -0.7 & 1093.33/819 \\
  &   & 2.04 $\pm$ 0.02 & 1.70 $\pm$ 0.04 & = 1.33 \\
\hline
3 & 8 & $11 \pm 5$ & -0.7 & 1212.85/819 \\
  &   & 1.96 $\pm$ 0.48 & 1.67 $\pm$ 0.06 & = 1.48  \\
\hline
4 & 8 & $11 \pm 5$ & -0.7 & 1241.95/819 \\
  &   & 1.76 $\pm$ 0.46 & 1.38 $\pm$ 0.06 & = 1.52  \\
\hline
5 & 8 & $11 \pm 5$ & -0.7 & 793.06/819 \\
  &   & 1.17 $\pm$ 0.48 & 0.94 $\pm$ 0.07 & = 0.97   \\
\hline
6 & 7.82 $\pm$ 0.98 & $11 \pm 5$ & -0.24 $\pm$ 0.27 & 995.85/817 \\
  &   & 1.20 $\pm$ 0.47 & 3.16 $\pm$ 0.2 &  = 1.22  \\
\hline
\end{tabular}
\end{center}
\end{table}

\begin{table}[tp]
\begin{center}
\caption{Fitting results of our atmospheric disk emission model 
to the \KU\ {\it Suzaku} data between 6-9 keV 
for the inclination angle $i=70\pm 6^{\circ} $ }
\label{fit:res2}

\begin{tabular}{*{5}{l}}
\multicolumn{5}{ c }{Energy range 6-9 keV} \\
\hline \hline
 Ep. & WABS & ATM & PL & $\chi^2/d.o.f.$ \\
  & $N_H$ ($10^{22}$) & Inc. (ang) & Ph. Index &   \\
  &   & Norm ($10^{-5}$) & Norm ($10^{-3}$) &   \\
\hline
1 & 8 & $70 \pm 6$ & -0.7 &  2236.77/819  \\
  &   & 4.78 $\pm$ 0.03 & 3.59 $\pm$ 0.03 &  = 2.73 \\
\hline
2 & 8 & $70 \pm 6$ & -0.7 & 1101.27/819 \\
  &   & 3.347 $\pm$ 0.032 & 1.79 $\pm$ 0.04 & = 1.34  \\
\hline
3 & 8 & $70 \pm 6$  & -0.7 & 1262.26/ 819 \\
  &   & 3.253 $\pm$ 0.024 & 1.76 $\pm$ 0.03 & = 1.54  \\
\hline
4 & 8 & $70 \pm 6$ & -0.7 &  1290.14/819 \\
  &   & 2.915 $\pm$ 0.023 & 1.46 $\pm$ 0.03 & = 1.57  \\
\hline
5 & 8 & $70 \pm 6$ & -0.7 & 796.61/819 \\
  &   & 1.957 $\pm$ 0.024 & 1.00 $\pm$ 0.03 & = 0.97  \\
\hline
6 & 7.37 $\pm$ 0.96 & $70 \pm 6$ & -0.7 $\pm$ 0.18 & 1033.43/817 \\
  &   & 2.035 $\pm$ 0.093 & 1.20 $\pm$ 0.08 & = 1.26  \\
\hline
\end{tabular}
\end{center}
\end{table}

Since back illuminated XIS1 chip improves the statistics only in low 
energy tail around 2 keV or less, it should cause no difference in the range 
between 6-9 keV. 
Therefore, for further analysis we take only spectra from front illuminated 
- XIS023 chips, as it is in the paper by \citet{kubota2007}.  

Choosing energy band between 6-9 keV we reduce number of degrees of freedom 
by a factor 2.5. We again fit {\sc wabs} * ({\sc atm} + {\sc powerlaw}) 
and results of this fitting are presented in Fig.~\ref{fig:band}. 
The importance of {\sc powerlaw} model is now lower 
than in case where fitting was done for full energy band 
(see Sec.~\ref{sec:cont}). The photon index is usually $-0.7 \pm 0.18$ 
and does not change much during different periods of observations. 
Similarly, column density of  galactic interstellar absorption 
$N_{H}$ in most cases equals $ 8.16 \pm 0.12 \times 10^{22}$ cm$^{-2}$. 
Since interstellar absorption should not change in different Epochs we 
fixed this parameter at $8 \times 10^{22}$  cm$^{-2}$.  
Only for observation number 6 we allow interstellar absorption to vary, but 
fit is still with good statistics. 
We show all fitting parameters in  Tables~\ref{fit:res}
and~\ref{fit:res2}.

Again, all six spectra imply the inclination angle $i= 11 \pm 5^{\circ}$, 
since absorption lines in the disk seen ``face on'' are the deepest in 
our model. Nevertheless, we have checked if the disk observed at the 
viewing angle $i=70^{\circ}$, appropriate for dipping sources can also 
explain observations. 
Surprisingly, fitting results with inclination frozen at $70^{\circ}$
also give quite good statistics as seen (see Tab.~\ref{fit:res2}). 
 
Most important results of this paper are presented in Fig.~\ref{fig:line},
which is enlarged part of Fig.~\ref{fig:band}. Resonant helium and 
hydrogen-like iron lines are clearly seen with their complex  profiles. 
Our absorption lines do not need any velocity shift, contrary to the previous
analysis \citep{kubota2007}. 
Absorption lines from the upper parts of static accretion disk atmosphere
can also satisfactorily explain observations. 

\section{Conclusions}
\label{sec:summary}

In this paper we presented fitting of complex continuum and line numerical 
models to X-ray spectra of \KU\ . In our models the spectrum of disk emission 
was obtained from careful radiative transfer computations including Compton 
scattering on free electrons. The Fe line profiles are computed 
as the convolution of natural, thermal and pressure broadening mechanisms. 
The advantage of our models 
is that the continuum is fitted with lines simultaneously, which was never done
before in the analysis of X-ray absorption lines seen in LMXBs. 
The usual procedure is to fit first the disk emission as a standard model 
in XSPEC fitting package, and Gaussian lines, where the energy of line 
centroid is a free parameter of fitting. In such the case, lines are usually 
blueshifted indicating that absorbing matter outflows. We show in this 
paper, that in the case of \KU\ there is sufficient to assume zero velocity
shift of absorbing matter. 
In our paper iron absorption lines originate in the hot atmosphere 
above accretion disk.

Our models are parametrized only by mass of the black hole, its spin
and disk accretion rate, which we assume to be constant at all radii. Since
our computations are very time consuming and we have only few collected
models, we are not able to derive those parameters from continuum fitting.  
On the other hand, quality of {\it Suzaku} data is high enough to follow the 
shape of continuum even in heavily obscured sources (in case of \KU\ 
obscuration due to Galactic absorption 
equals ca. $8\times 10^{22}$ cm$^{-2}$ ).
 
Accretion disk atmosphere spectra fits {\it Suzaku} data for \KU\ very well. 
The best fit we have obtained for the inclination angle $i=11^{\circ}$. 
For higher angles i.e. $i=70^{\circ}$ fit is just slightly worser. 
This angle is within the range of inclination suggested in the literature 
taking into account dipping behavior and assuming absorption in the wind. 
Small difference of fit quality between different inclinations does not 
allow us to claim constraints on $i$.

Of course models such as {\sc diskbb+pl} plus absorption 
with two Gaussians
always show better statistics than our more complex model. But 
the former  model has many free parameters. This is a general rule that 
the more free parameters we have the more accurate is the final fit. 
The our model has no free parameters connected with lines, therefore, 
it has at least 6 parameters less than the set {\sc diskbb+pl} plus 
two Gaussians 
applied by \citet{kubota2007}.

According to any global disk models considered for the mass of 
central object of the  order 10  $M_{\odot}$ or less, the effective
temperature of the inner radii reaches $10^7$K. In such a case 
possible origin of absorption lines from accretion disk atmospheres should
be taken into account in modeling of winds in X-ray binaries.

In the wind theory, it is widely accepted that the wind can be launched 
at the accretion disk surface. Upper layers of atmosphere can become 
unstable and start to blow out material due to radiation pressure. 
Our analysis does not contradict this fact, it rather shows that absorption 
at upper atmospheric layers cannot be distinguished from the absorption 
in the wind, which may be launched in the same region. 

In this work we do not aim to tightly constrain parameters of the object 
but rather show that emission from the accretion disk atmosphere is an 
important mechanism which gives vital explanation or at least part 
of an answer to the question of origin of iron absorption in X-ray binaries. 

The major conclusion of our analysis is that the shape of disk spectrum 
is well interpreted as \KU\ emission and that absorption lines do not need
to set any velocity shift to explain data. Therefore, the wind explanation
for absorbing matter is questionable and not unique. We showed that X-ray
data of current quality can be interpreted in several ways and we cannot
easily solve this ambiguity.
We conclude that the wind theory can be an artifact of the fitting 
procedure, when 
the continuum and lines are fitted as separate model components. 
Data of higher spectral resolution are needed to discriminate between two 
models. Future satellites with calorimeters as {\it ASTRO-H} or {\it Athena+}
will yield us the answer.

%
\begin{acknowledgements}
This research was supported by the Polish Ministry of Science 
and Higher Education grant No. N N203 511638, and by  Polish National Science 
Center grant No. 2011/03/B/ST9/03281, and has received funding
from the European Union Seventh Framework Program (FP7/2007-2013) under 
grant agreement No.312789.
\end{acknowledgements}

\bibliographystyle{aa}
\bibliography{refs}

\end{document}